\documentclass[doublecol]{epl2} 

\title{Strong temperature dependence of antiferromagnetic coupling in CoFeB/Ru/CoFeB}

\author{N. Wiese\inst{1,2}\thanks{\emph{Present address:} University of Glasgow, Kelvin Building, Glasgow G12 8QQ, United Kingdom, \email{n.wiese@physics.gla.ac.uk}} \and T. Dimopoulos\inst{1}\thanks{\emph{Present address:} ARC Nano-System-Technologies, Donau-City-Str. 1, 1220 Vienna, Austria} \and M. R\"uhrig\inst{1} \and J. Wecker\inst{1} \and G. Reiss\inst{2} \and J. Sort\inst{3} \and J. Nogu\'{e}s\inst{3}
}

\shortauthor{N. Wiese \etal}

\institute{
  \inst{1}Siemens AG Corporate Technology, Paul-Gossen-Stra{\ss}e 100, 91052 Erlangen, Germany \\
  \inst{2}University of Bielefeld, Nano Device Group, Universit\"atsstra{\ss}e 25, 33615 Bielefeld, Germany\\
  \inst{3}Instituci\'{o} Catalana de Recerca i Estudis Avan\c{c}ats (ICREA) and Departament de F\'{i}sica, Universitat Aut\`{o}noma de Barcelona, 08193 Bellaterra, Spain
}


\abstract{
The temperature dependence of saturation and spin-flop fields for artificial ferrimagnets (AFi) based on antiparallel coupled CoFeB/Ru/CoFeB trilayers has been investigated in a temperature range between 80K and 600K. The results presented in this paper are relevant for magnetic devices using this system, e.g. magnetic-random access memory based on spin-flop switching. In good accordance to the theory, the saturation field $H_{\mbox{\scriptsize sat}}$ behaves like $H_{\mbox{\scriptsize sat}} \propto H_0 (T/T_0)/\sinh(T/T_0)$ with a characteristic temperature of $T_0 \approx 150$K. Within this model, the Fermi velocity for the Ru layer is of the order of $10^{5}$m/s, therefore, explaining the strong variation of the coupling strength with the temperature in Ru based AFi. Furthermore, a strong uniaxial anisotropy of $K_u = 2\times 10^3$J/m$^3$ with a small angular distribution of the anisotropy axes is observed for the AFi trilayers based on amorphous CoFeB alloys.
}

\begin{document}

\maketitle
\section{Introduction}
\label{intro}
Magnetic tunnel junctions (MTJ) have gained considerable interest in recent years due to their high potential as sensor elements \cite{Berg99} and in magnetic random access memories (MRAM).\cite{Gallagher97} The basic design of a spin valve consists of a hard magnetic reference electrode separated from the soft magnetic sense or storage layer by a tunnel barrier like $Al_2O_3$.

Recently, a technique known as {\em spin-flop switching} has been announced for the use in MRAM, using artificial ferrimagnets (AFi) as the soft magnetic electrode and the use of bit toggling rather than switching.\cite{Engel05} In this toggling scheme the spin-flop or plateau field, $H_{\mbox{\scriptsize p}}$, has to be overcome, thus inducing a small net moment of the magnetizations of the AFi electrode. The bit is then reversed by a rotation of the net moment.

\begin{figure}
\begin{center}
\includegraphics[width=5.5cm]{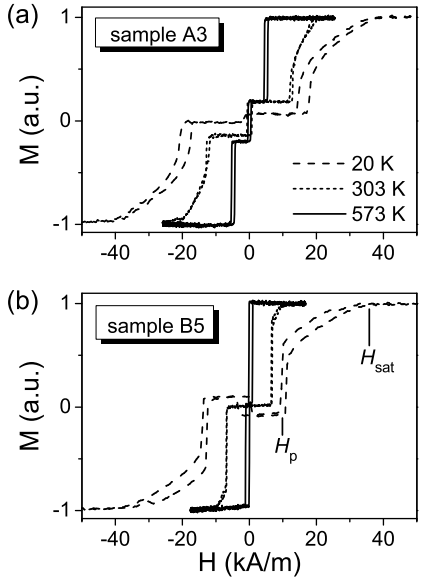}
\caption{\label{figure1} Magnetization loops at various
temperatures, $M(H,T)$, obtained by MOKE for (a) a sample of
CoFeB(4.5nm)/Ru(0.95nm)/CoFeB(3nm) (sample A3) and (b)
CoFeB(3.5nm)/Ru(1.05nm)/CoFeB(3nm) (sample B5). The saturation field, $H_{\mbox{\scriptsize sat}}$, and the plateau field, $H_{\mbox{\scriptsize p}}$, are indicated for one of the loops of sample B5.}
\end{center}
\end{figure}

It has been experimentally shown, that the exchange coupling strength can significantly vary with temperature.\cite{Celinski90} Within the quantum-well model of coupling, the coupling strength, $J = - \mu_0 H_{\mbox{\scriptsize sat}} (m_1 m_2)/(m_1+m_2)$ \cite{Berg97}, where $m_{1,2}$ are the magnetization of the two ferromagnetic layers, shows a strong dependence on temperature $T$, and follows the relationship
\cite{Edwards91,Bruno91}
\begin{equation}
J(T) = J_0 \frac{(T/T_0)}{\sinh (T/T_0)}
\label{coupling_temperature}
\end{equation}

The characteristic temperature is given by $T_0 = \hbar v_{F} / (2 \pi k_B t_{\mbox{\scriptsize NM}})$, where $v_F$ is the Fermi velocity and $t_{\mbox{\scriptsize NM}}$ the thickness of the nonmagnetic spacer layer. This dependence was first experimentally confirmed by Zhang et al. in Co/Ru/Co trilayers.\cite{Zhang94} In those studies the characteristic temperature for Ru was of the order of $100$K, thus resulting in a Fermi velocity of ${v_F \approx 10^{5}}$m/s in Ru. This value is approximately one order of magnitude lower than for most nonmagnetic metals, which gives rise to a strong temperature dependence of the coupling strength in Ru-based AFi systems.\cite{Persat97}

It was the purpose of this study, to investigate the saturation (and, therefore, the coupling strength) and the plateau field in CoFeB based AFi systems as a function of the temperature. The results are obtained from temperature dependent magnetization loops, $M(H,T)$, taken over a wide temperature range, using unpatterned CoFeB/Ru/CoFeB AFi systems. The Co$_{60}$Fe$_{20}$B$_{20}$ alloy was chosen for the high TMR ratios up to 70$\%$ achieved in combination with AlO$_x$ based barriers,\cite{Wang04} or even up to $472\%$ in MgO based MTJs.\cite{Hayakawa06} The CoFeB/Ru/CoFeB AFi shows a low antiferromagnetic (AF) coupling strength in the order of $-0.1$mJ/m$^2$ in the 2nd AF coupling maximum and has been successfully integrated into a complete MTJ stack.\cite{Wiese04} Furthermore, the thermal stability of these systems has been proven for annealing temperatures up to $\sim 350^{\circ}$C.\cite{Wiese05} Therefore, this material system is a promising candidate for integration as a soft magnetic electrode in MTJs.

\section{Experiment}
All samples have been deposited by RF and DC sputtering on thermally oxidized SiO$_2$ wafers at a base pressure of $5 \times 10^{-8}$mbar. A magnetic field of approximately 4 kA/m was applied during deposition in order to predefine the easy axis in the magnetic layers. The AFi was grown on a 1.2nm thick Al layer, oxidized in an Ar/O$_2$ plasma for 0.8 min without breaking the vacuum, to reproduce the growth conditions in a full MTJ stack. All samples have been protected from oxidation by a Ta(5nm)/TaN(5nm) capping layer.

\begin{figure}
\begin{center}
\includegraphics[width=8.6cm]{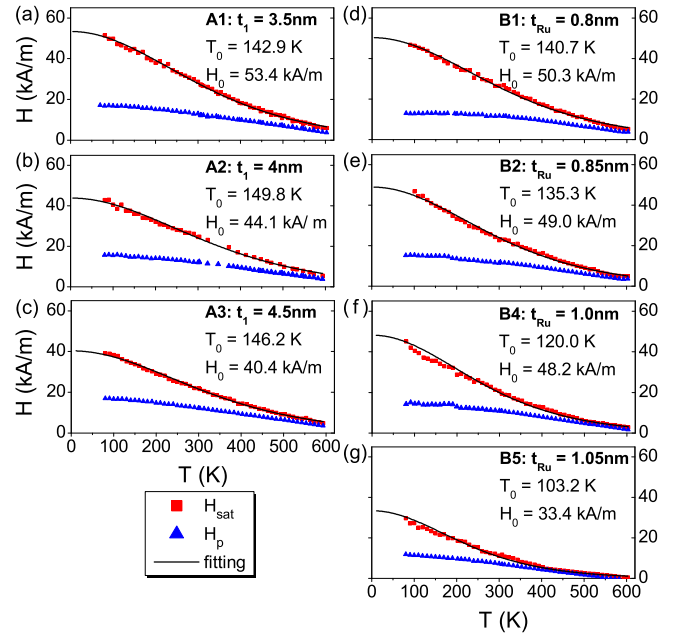}
\caption{(Color online) Saturation and plateau fields with varying
temperature for (a)-(c) samples of series A and (d)-(g) series B,
respectively. The lines are fits of the $(T/T_0)/\sinh(T/T_0)$
dependence to the experimental data. The extracted values of the
zero temperature saturation field, $H_{0}$, and the characteristic
temperature, $T_0$, are given at the plots.} \label{figure2}
\end{center}
\end{figure}

Two series of samples have been prepared. In series A, the thickness $t_1$ of the magnetic layer interfacing with the AlO$_x$ was varied from 3.5 to 4.5nm in steps of 0.5nm (CoFeB ($t_1$)/Ru(0.95nm)/CoFeB(3nm)). In series B the Ru thickness, $t_{\mbox{\scriptsize Ru}}$, was varied from 0.8 to 1.05nm around the 2nd AF coupling maximum, while using an almost compensated CoFeB(3.5nm)/Ru($t_{\mbox{\scriptsize Ru}}$)/CoFeB(3nm) AFi (see fig. \ref{figure2} for nomenclature). After deposition, all samples have been heat treated at 300$^{\circ}$C for 10min. and cooled down to RT under application of a constant magnetic field of $\sim 40$kA/m.

For the measurement of magnetization loops versus temperature, two different setups have been used, depending on the temperature range. For $80$K$<T<300$K a cryostat was placed between the pole gaps of our magnetooptical Kerr effect (MOKE) magnetometer, whereas for $T>300$K a simple heating stage mounted in air has been used.\cite{NanoMOKE} The measurements have been performed using a focused laser with a spot diameter of $\sim 4\mu$m at a field sweeping frequency between 3Hz and 12Hz. All together, magnetization loops, $M(H,T)$, in a temperature range from 80K up to 600K have been recorded in steps of 10K.

Before performing the MOKE measurements with varying temperature, M(H) loops of all samples have been obtained by alternating gradient field magnetometery (AGM) at room temperature (RT).


\section{Results and Discussion}
 From the AGM measurements we extract the saturation field, $H_{\mbox{\scriptsize sat}}$, and the net and total magnetic moment, thus allowing to calculate the absolute value of the coupling strength at room temperature, $J^{\mbox{\scriptsize AGM}}$. The coupling strength has a constant value of $-0.06$mJ/m$^2$ for the samples of series A, whereas for series B the coupling strength varies with spacer thickness from $-0.02$ to $-0.06$mJ/m$^2$ with the maximum for $t_{\mbox{\scriptsize Ru}}=0.95$nm in accordance with the 2nd AF maximum reported previously.\cite{Wiese04}

Temperature dependent magnetization loops obtained by MOKE, $M(H,T)$, are shown in fig. \ref{figure1} for two samples of series A and B, respectively. The loops exhibit a typical AFi behavior. The hysteresis at low fields arises from the non-compensation of the two AF coupled ferromagnetic (FM) layers, and is attributed to a switching of the net moment of the antiparallel aligned layers. The large magnetization change at higher fields take place when the layers flop from an antiparallel to a parallel arrangement of the magnetization. Loops of all samples have been recorded, and at RT they show identical behavior as the AGM measurements discussed before. Moreover, it can be observed that both, $H_{\mbox{\scriptsize sat}}$ and $H_{\mbox{\scriptsize p}}$, decrease with increasing temperature. In fig. \ref{figure2}, the temperature dependence of $H_{\mbox{\scriptsize sat}}(T)$ and $H_{\mbox{\scriptsize p}}(T)$ is shown for all the samples under study. The saturation field data have been fitted with eq. \ref{coupling_temperature}. The experimental data show a good agreement with the fitted model, and one can obtain the values of the zero temperature saturation field, $H_0$, and the characteristic temperature, $T_0$.

Since the Ru spacer thickness and the coupling has been held constant for the samples of series A, $T_0$, and accordingly, $v_F$, are almost constant. For the samples of series A, the extracted values for $T_0$ are approximately $145$K. This yields to a Fermi velocity of ${v_F \approx 1.15 \times 10^{5}}$m/s, which is in accordance with the values obtained in Co/Ru/Co multilayers.\cite{Zhang94}


\begin{figure}
\begin{center}
\includegraphics[width=5.5cm]{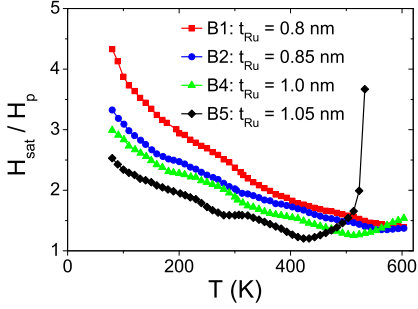}
\caption{(Color online) Temperature dependence of $H_{\mbox{\scriptsize sat}}/H_{\mbox{\scriptsize p}}$ for samples with varying $t_{\mbox{\scriptsize Ru}}$ (series B). All curves have been smoothed by averaging, using two adjacent datapoints.} \label{figure3}
\end{center}
\end{figure}

Secondly, it is observed from the $M(H,T)$ measurements, that $H_{\mbox{\scriptsize sat}}$ depends more strongly on the temperature than $H_{\mbox{\scriptsize p}}$ in the samples under study. Therefore, the ratio $H_{\mbox{\scriptsize sat}}/H_{\mbox{\scriptsize p}}$ decreases for all samples with increasing temperature (see fig. \ref{figure3}). The separation of plateau and saturation field, necessary for the spin-flop writing of MRAM bits, is significantly reduced for elevated temperatures. In terms of the magnetic phase diagram, calculated for AFi structures by Worledge in ref. \cite{Worledge04}, all samples get closer to the limit between the spin-flop and the metamagnetic phase, defined by the condition $H_{\mbox{\scriptsize sat}}=H_{\mbox{\scriptsize p}}$, with increasing temperature. Furthermore, samples with $t_{\mbox{\scriptsize Ru}} \ge 1.0$nm show a vanishing plateau field at elevated temperatures, causing a strong increase of $H_{\mbox{\scriptsize sat}}/H_{\mbox{\scriptsize p}}$ for $T>510$K and T$>425$K, respectively. This behavior most likely indicates, that the weak AF coupling for these samples is overcome by thermal activation. However, for the experimental conditions in this study, the process was reversible with decreasing temperature.

\begin{figure}
\begin{center}
\includegraphics[width=8.6cm]{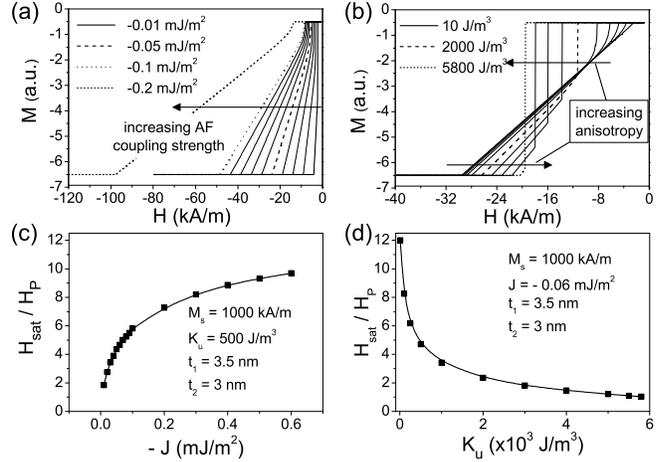}
\caption{\label{figure4} Simulated magnetization loops in dependence
on (a) the AF coupling strength, $-J$, and (b) the uniaxial
anisotropy, $K_{\mbox{\scriptsize u}}$. Values of
$H_{\mbox{\scriptsize sat}}/H_{\mbox{\scriptsize p}}$ extracted from
the M(H) loops for (c) varying $-J$ and (d) $K_{\mbox{\scriptsize
u}}$. The simulation parameters used are denoted.}
\end{center}
\end{figure}

To investigate the origin of this behavior, the magnetization loops
of the AFi system have been simulated, using a simple minimization
of the total energy of the AFi. The model considers the individual
magnetic moments of the FM layers, the bilinear coupling energy, the
uniaxial anisotropy energy, and the Zeeman energy of the applied
external magnetic field. The resulting $M(H)$ behavior for one
branch of the magnetization loops are shown for different values of the AF
coupling strength and the uniaxial anisotropy constant in fig.
\ref{figure4}(a) and (b), respectively. The ratio
$H_{\mbox{\scriptsize sat}}/H_{\mbox{\scriptsize p}}$ has been
evaluated from these loops and is shown in fig. \ref{figure4}(c)
and (d). It is evident, that the field range between spin-flop and
saturation field can be increased, if either the AF coupling is
increased or the intrinsic anisotropy of the AFi is decreased.

Comparing the experimental and simulated M(H) loops, one can conclude that all samples show a very strong anisotropy at RT. If the coupling strength and the magnetic moments evaluated from the AGM measurements are taken into account, the experimental data are best fitted with an uniaxial anisotropy constant $K_{\mbox{\scriptsize u}} = 2000$J/m$^3$. Most interestingly, the experimental data show a pronounced plateau and an abrupt change of the magnetization for $H > H_{\mbox{\scriptsize p}}$. This behavior is most likely attributed to a small angular distribution of the anisotropy axes, probably associated with the amorphous nature of the layers, and is not observed in comparable stacks of polycrystalline materials. While the narrow distribution is essential for application of these AFis as soft magnetic electrode in MTJs, it can be inferred from the simulations and the experimental data that the absolute value of the anisotropy has to be decreased. This would increase the field range between plateau and saturation field, necessary for spin-flop switching, especially important for operation at elevated temperatures or in heat assisted writing schemes.

\begin{figure}
\begin{center}
\includegraphics[width=6.5cm]{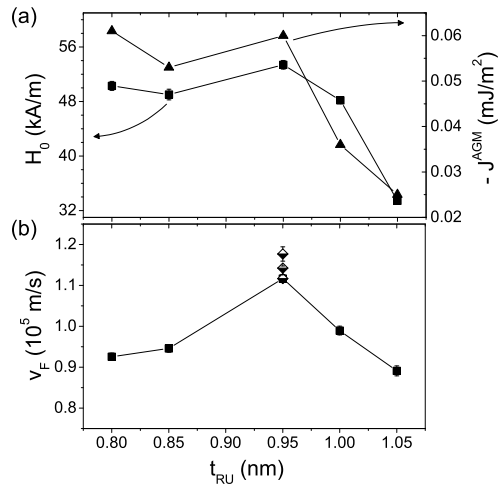}
\caption{\label{figure5} (a) Dependence of the room temperature coupling strength ($J^{\mbox{\scriptsize AGM}}$), obtained from AGM measurements, and zero temperature saturation field ($H_0$) on the Ru spacer thickness. (b) Evaluated Fermi velocities in dependence on the Ru spacer thickness. The three data points at $t_{\mbox{\scriptsize Ru}} = 0.95$nm are from the samples of series A, showing an almost independent Fermi velocity on the variation of the net moment.}
\end{center}
\end{figure}

From the quantum well model of coupling it is predicted, that the exchange coupling oscillates in sign with a period of $\pi / k_{\mbox{\scriptsize F}}$, and the amplitude of the oscillation decays like $1/t_{\mbox{\scriptsize NM}}$, where $k_{\mbox{\scriptsize F}}=\frac{\hbar v_{\mbox{\scriptsize F}}}{m_{\mbox{\scriptsize e}}}$ is the Fermi wave vector in the spacer layer.\cite{Stiles02} In fig. \ref{figure5}(a) the dependence of the coupling strength, $J^{\mbox{\scriptsize AGM}}$, and the zero temperature saturation field, $H_0$, on the Ru thickness are presented. Both values show a maximum at $t_{\mbox{\scriptsize Ru}}=0.95$nm, in good accordance to the position of the 2nd AF coupling maximum as previously reported.\cite{Wiese04} In fig. \ref{figure5}(b) the evaluated dependence of $v_{\mbox{\scriptsize F}}$ on the Ru spacer thickness is shown, and a maximum of Fermi velocitiy around the second AF coupling maximum at $t_{\mbox{\scriptsize Ru}}=0.95$nm is found.

\section{Conclusion}
The temperature dependence of the saturation field, and therefore the coupling strength, for CoFeB/Ru/CoFeB trilayers has been investigated. From these studies, one can evaluate the characteristic
temperature to be about 150K and the inferred Fermi velocity of $1.15 \times 10^{5}$m/s, in good accordance with results for Co/Ru/Co trilayers in the literature. The AFis based on amorphous CoFeB show a high anisotropy with a small angular distribution of the anisotropy axes. From the experimental results and simulations it is shown that a reduction of the anisotropy would be desirable to improve the high temperature performance of these AFis. 

\acknowledgments
The authors wish to thank G. Gieres and H. Mai for discussions and experimental support. J.N and J.S. acknowledge the partial financial support from the Institut Catal\`{a} de Nanotecnologia (ICN), and the 2005SGR-00401 and MAT-2004-01679 research projects.

\end{document}